\documentclass[letterpaper]{article} % DO NOT CHANGE THIS
\usepackage{aaai25}  % DO NOT CHANGE THIS
\usepackage{times}  % DO NOT CHANGE THIS
\usepackage{helvet}  % DO NOT CHANGE THIS
\usepackage{courier}  % DO NOT CHANGE THIS
\usepackage[hyphens]{url}  % DO NOT CHANGE THIS
\usepackage{graphicx} % DO NOT CHANGE THIS
\usepackage{natbib}  % DO NOT CHANGE THIS AND DO NOT ADD ANY OPTIONS TO IT
\usepackage{caption} % DO NOT CHANGE THIS AND DO NOT ADD ANY OPTIONS TO IT
\usepackage{algorithm}
\usepackage{algorithmic}
\usepackage{enumitem}

\usepackage{xcolor}
\newcommand{\answerYes}[1]{\textcolor{blue}{#1}} 
 
\newcommand{\answerNA}[1]{\textcolor{gray}{#1}}

\usepackage{newfloat}
\usepackage{listings}
\DeclareCaptionStyle{ruled}{labelfont=normalfont,labelsep=colon,strut=off} % DO NOT CHANGE THIS
\floatstyle{ruled}
\newfloat{listing}{tb}{lst}{}
\floatname{listing}{Listing}
\title{Mapping the Climate Change Landscape on TikTok}

\author{
    Alessia Galdeman,
    Luca Maria Aiello
}
\affiliations{
    IT University of Copenhagen
    \\
    
    gald@itu.dk, luai@itu.dk
}

\begin{document}

\maketitle

\begin{abstract}

Social media platforms shape climate action discourse. Mapping these online conversations is essential for effective communication strategies. TikTok's climate discussions are particularly relevant given its young, climate-concerned audience. In this work, we collect the first TikTok dataset on climate topics. We collected 590K videos from 14K creators along with their follower networks. By applying topic modeling to the video descriptions, we map the topics discussed on the platform on a climate taxonomy that we construct by consolidating existing categorizations. Results show TikTok creators primarily approach climate through the angle of lifestyle and dietary choices. By examining semantic connections between topics, we identified non-climate "gateway" topics that could draw new audiences into climate discussions.
\end{abstract}

\section{Introduction}

Social media platforms are vital spaces for global climate change dialogue~\cite{pearce2019social}. While enabling broader participation, understanding their impact requires analyzing emerging platforms and engagement patterns. Traditional research has primarily focused on text-based platforms such as Twitter and Reddit~\cite{fownes2018twitter, effrosynidis2022climate, treen2022discussion}, but contemporary social media landscapes are increasingly dominated by short-form video content, particularly among younger audiences~\cite{hautea2021showing}.
Content creators on these platforms serve as influential information disseminators and opinion leaders, wielding significant power in shaping climate discourse and potentially bridging polarized viewpoints~\cite{falkenberg2022growing}. However, the characteristics and dynamics of modern platforms like TikTok remain relatively unexplored in the literature~\cite{corso2024we}, particularly regarding how broader topics are related to the climate change discussions. 
While previous studies have examined climate content on YouTube and its role in activist communities~\cite{shapiro2018climate, uldam2013online}, research on TikTok has been limited to content analysis and linguistic features~\cite{hautea2021showing, basch2022climate, nguyen2023tiktok} or to small-scale quantitative studies on climate video creators~\cite{pera2024shifting}. As a result, there remains a crucial gap in understanding how climate discussions are shaped across modern platforms and identifying topics that act as bridges towards climate-related topics.
This work addresses these gaps through two main contributions. 
First, we create a dataset containing $590361$ climate-related TikTok videos and the social network between their $14K$ creators. We map these videos onto a taxonomy of climate topics that we obtain by combining existing literary sources. Second, we explore how different topics are related to one another, identifying non-climate topics that are semantically close to the climate discussion, thus having the potential to introduce new audiences to the climate discourse. We addressed two main research questions:

\vspace{1mm} \noindent \textbf{RQ1:} \textit{What are the climate topics discussed on TikTok and in what proportion? Do they cover known taxonomies?}

\vspace{1mm} \noindent \textbf{RQ2:} \textit{What non-climate topics on TikTok are potential `gateway' topics for people to start engaging with climate topics?}

\section{Taxonomy of Climate Topics} \label{sub:taxonomy}
\begin{figure}[t!]
    \centering
    \includegraphics[width=\linewidth]{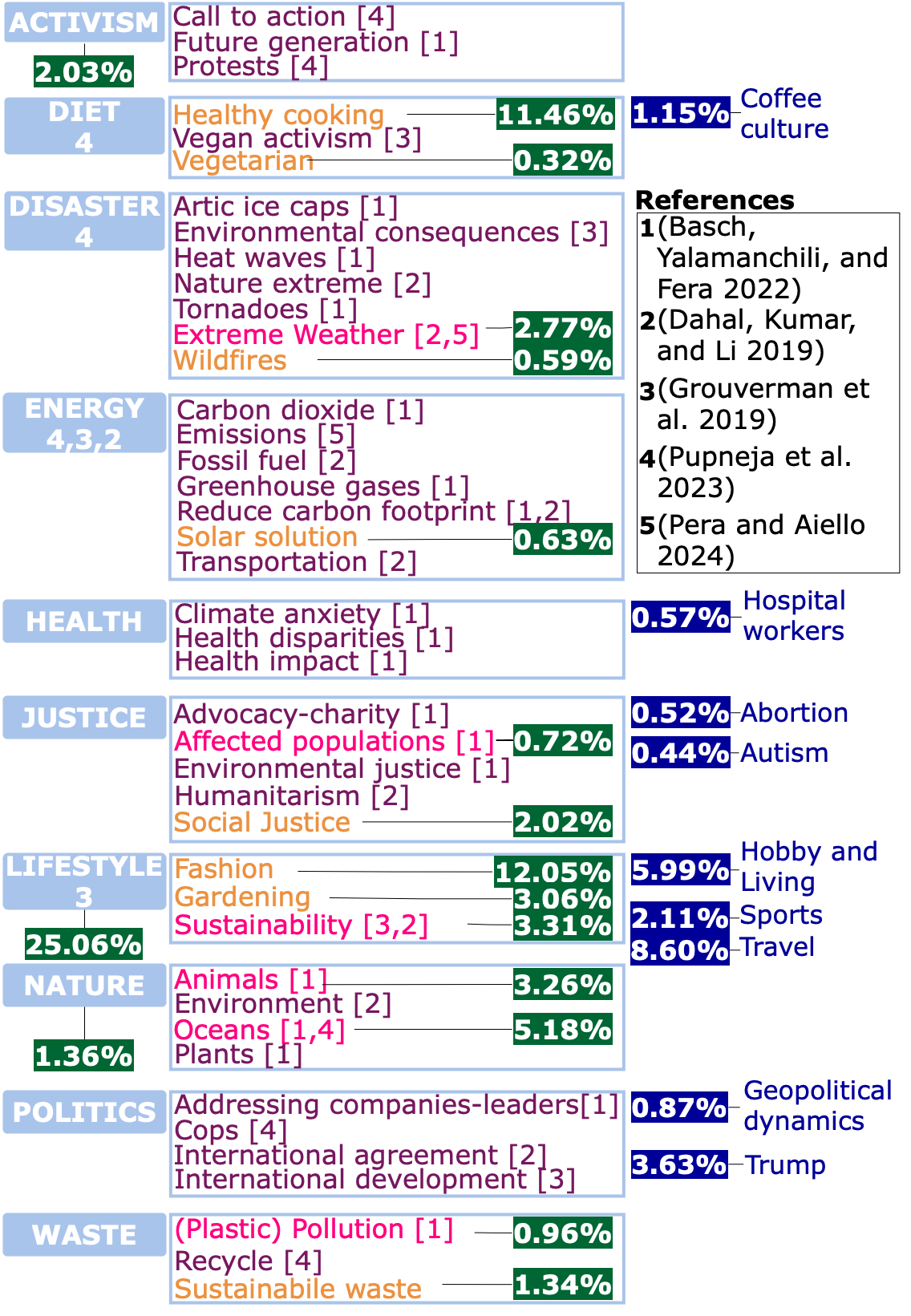}
    \caption{\textbf{Taxonomy of climate-related topics.} Labels are color-coded based on their occurrence: purple indicates presence only in previous works, orange shows topics found exclusively in our data collection, and magenta represents topics present in both. Non-climate topics are shown in blue. Superscript numbers on purple and magenta topics indicate the work where these topics were originally identified. }
    \label{fig:taxonomy}
\end{figure}

To create a taxonomy of climate change topics, we aggregated categorizations proposed by existing studies~\cite{basch2022climate,dahal2019topic,grouverman2019climate,pera2024shifting,pupneja2023understanding}. These are empirical studies on either Twitter or TikTok. They all provide topic lists extracted from their specific video collections, which often focus on particular subtopics and therefore do not exhaustively capture all the facets of the climate change discourse. 
To the best of our knowledge, this work represents the first attempt to create a data-driven taxonomy of climate change topics on TikTok, informed by both empirical evidence and existing literature. We first collected all topic labels from the referenced studies (51 in total). We then manually merged near-duplicates and grouped the labels into 10 semantically coherent top-level categories. The resulting taxonomy is visualized in Figure~\ref{fig:taxonomy}.

\section{Data Collection} \label{sec:data}

We collect all the data through the TikTok Research API \cite{TikTokDevelopers}, and make the dataset public\footnote{Dataset's DOI: \texttt{10.5281/zenodo.15131067}}.
First, we identify the set of hashtags that are used on TikTok to tag climate-related content. To do so, we start by filtering 6 existing lists \cite{greenpeaceClimateVoices, wiredMeetClimate, palauprojectEnvironmentalInfluencers, nytimesWithTikTok, climatecreatorstowatchClimateCreators, ecoktokcollective} of climate-change centered accounts. From these lists, we obtain $49$ \textit{seed users}, highly active on TikTok and specialized in posting videos related to climate change. We deliberately exclude news outlets from this list as they are not climate change-focused accounts. To mitigate the potential bias of this initial selection, we augment the seed set with a snowball expansion \cite{goodman1961snowball}. From the $16,533$ accounts followed by our seed users, we selected the top $0.5\%$ based on the number of follows from seed users. We then manually reviewed these accounts for topical relevance, adding the climate-dedicated profiles to our seed set.
This step yields $56$ new seed users, for a total of $105$. We extract the metadata of all $17,338$ public videos posted by seed users, capturing their complete posting history from their first video on the platform until the time of data collection (October 2024). We manually review the relevant hashtags in the $99^{th}$ percentile of the frequency distribution and excluded irrelevant hashtags such as engagement tags (e.g., \texttt{$\#$viral}). Since our aim is to capture also the broader topical context around climate change discussions, we keep both climate-related keywords, as well as broader contextual hashtags that are still relevant to our taxonomical categories (e.g., \texttt{$\#$politics}). The final list contains $193$ hashtags, which we manually assign to the most fitting taxonomical branches (Table~\ref{tab:topicsize} in Appendix).

We query the API to collect videos for each hashtag separately. To retrieve a comparable number of videos across the ten taxonomical categories, we define a per-hashtag video quota $q_h$ that is inversely proportional to the number of hashtags in its category, namely: $q_h = \frac{37*100}{T(h)}$ where $T(h)$ is the number of hashtags in the taxonomical category of hashtag $h$ and 37 is the maximum value of $T(h)$. We collected videos in two-week intervals for each hashtag, from 16 October 2021 to 15 October 2022. This timeframe encompasses significant climate-related events including COP26, major climate protests, and extreme weather events. In total, we collected 1,326,684 videos from 199,968 users.

Finally, we construct the follower network among users. We first restrict our sample to the $32,765$ users with a minimum activity level of two videos posted within the timeframe of interest. We then collect the $536,585$ follower connections outgoing from these creators, and retained the links whose both endpoints were in our user set. After discarding singleton nodes, we were left with a network of $13867$ \emph{creators} connected by $161,891$ links.

\section{Climate Topics on TikTok} \label{sec:topic_modeling}

Hashtags are only broadly indicative of a video's theme. To obtain a more detailed description of the topics discussed in the videos we gather, we augment the video metadata with the video descriptions.
To do that, we focus on the $590361$ video English-language descriptions longer than three words after the removal of stopwords and non-English words. For each video, we create a document by concatenating its description text and hashtags. To these documents, we apply a BERTopic pipeline~\cite{grootendorst2022bertopic} that i) generates sentence-BERT embeddings for each document~\cite{reimers2019sentencebert}; ii) reduces the embedding dimensionality with UMAP~\cite{mcinnes2018umap}; and applies HDBSCAN clustering to group descriptions into topics~\cite{campello201hdbscan}. The optimal parameters for this pipeline were determined through Randomized Cross-Validation search, maximizing the relative validity measure \cite{moulavi2014density}. The hyperparameters corresponding to the best combination are: \texttt{min\_cluster\_size} $= 1000$, \texttt{min\_samples} $= 500$,  \texttt{cluster\_selection\_epsilon} $= 0.05$.
This resulted in 45 distinct clusters.
We label the topics using the support of a Large Language Model (LLM). We randomly sample $20\%$ of the documents from each cluster and parse each of them with the \texttt{gpt-4o-mini} model (OpenAI API \cite{OpenAmodels}) to assign a single short video label. In a second stage, we ask the same model to consolidate the individual topics into a single, coherent cluster label. Both the individual document labeling and the consolidation steps were performed independently on all 45 clusters (prompts in Appendix). Finally, we manually checked and grouped similar topics whose labels have high semantic similarity (e.g., \emph{Sustainable fashion} and \emph{Thrift fashion} under \emph{Fashion}), which results in 26 topics, 17 of which are related to climate.

In Figure \ref{fig:taxonomy}, climate-related subtopics, along with their volume percentage, are shown in one of the three colors: \textit{(i)} purple, if they appear only from a theoretical point of view; \textit{(ii)} orange, if they appear only in our data collection; \textit{(iii)} magenta, if they appear in both cases. Non-climate topics are colored in blue and positioned around their parent topic to complete the taxonomical structure. TikTok content covers only a few taxonomical categories, with most videos focusing on climate topics related to either \emph{Lifestyle}, \emph{Diet}, or \emph{Nature}.

\section{Boundaries of Climate Discussion}

We explore the interface between climate and non-climate content on TikTok, with the objective of charting potential gateway topics to introduce new users to the climate discussion. To do so, we use two complementary analytical perspectives. The first estimates the similarity between topics by leveraging the co-occurrence of topics in the video production of TikTok creators (\emph{topic perspective}). The latter looks at follower links from creators who do not typically discuss climate-related topics to climate creators (\emph{user perspective}).

\begin{figure}
    \centering
    \includegraphics[width=0.99\linewidth]{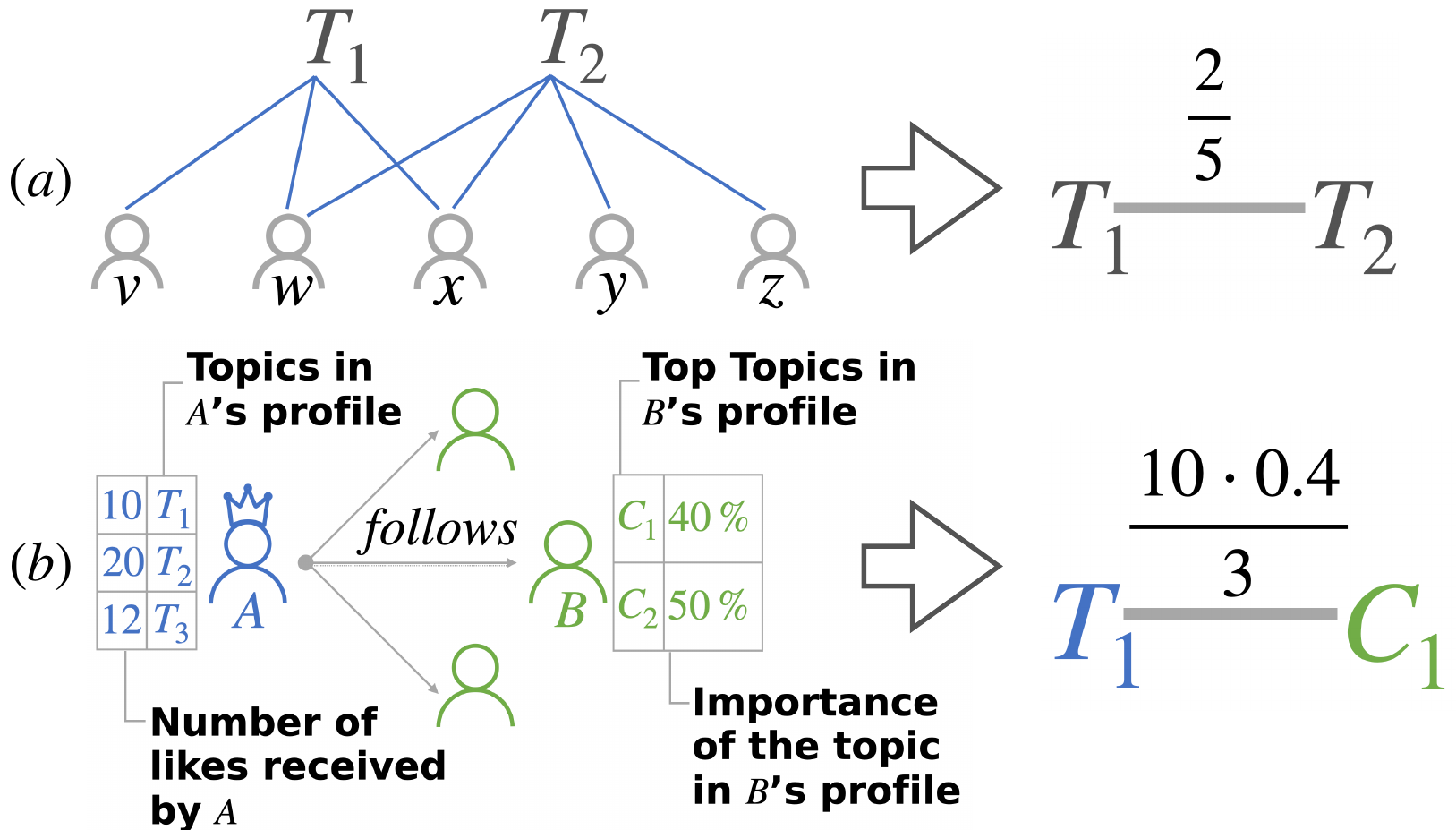}
    \caption{(a) \textbf{Projection of the user-topic bipartite network.}  Topic $T_1$ is discussed with high frequency by 3 users ($v,w,z$), while topic $T_2$ by $4$ users ($w,x,y,z$). In the projected graph on the topic layer, the edge $(T_1, T_2)$ has $w = \frac{2}{5}$. (b) \textbf{Climate and non-climate bridge topic pairs.} Non-climate user $A$ has received $10$ likes on topic $T_1$. This is divided by its 3 neighbors and weighted by $40\%$ as it represents the strength of topic $C_1$ for climate-user $B$. So, the weight of the edge $(T_1, C_1)$ concerning the couple of users $(A, B)$ is $\frac{10\cdot 0.4}{3}$. }
    \label{fig:q2schema}
\end{figure}

To implement \textbf{topic perspective}, we construct a bipartite network that connects users to their respective topics. An edge from user $u$ to topic $t$ has a weight $w$ representing the proportion of the user's video production dedicated to that particular topic. To focus on strong topic associations, we retain only links with weights in the top quartile of the distribution. We used the Jaccard index to project this bipartite network onto the topic dimension. The resulting network is an undirected graph where the edge connecting topics $T_1$ and $T_2$ is weighted by the ratio of the number of creators who have uploaded videos in both $T_1$ and $T_2$ over the number of creators who have uploaded videos on either $T_1$ or $T_2$ (Figure~\ref{fig:q2schema}a). 

To implement \textbf{user perspective}, we first profile creators to distinguish climate creators from non-climate creators, and then look at the following links between them. We characterize creators with a 26-dimensional profile vector that encodes the fraction of videos they published on each topic. We then compute a \textit{greenness} score for each user, defined as the percentage of their published videos that belong to climate-related topics. We use the follow network between creators and our established greenness metric to evaluate how climate-related videos might gain new viewers through the recommendations of influential creators whose production is not focused on climate, but who follow climate-related creators that they could advertise. We stick to a conservative scenario where we look at follower links from creators with no video on climate ($0\%$ \emph{greenness}) to those with full focus on climate topics ($100\%$ \emph{greenness}). We consider only non-climate top influencers above the 75th percentile of the distribution of number of likes received.

For each edge connecting a non-climate node to a climate node, we create the respective topic pairs, selecting all non-climate topics for the first node and the most discussed topics for the climate node (i.e., covering at least $62.5\%$ of the user profile, corresponding to the $90$th percentile of the coverage distribution). For a non-climate node $A$ and climate node $B$, the weight of a topic pair $(T_1, C_1)$ is calculated as the number of likes that $A$ received on topic $T_1$ (a proxy of $A$'s audience reach on topic $T_1$), divided by $A$'s out-degree (i.e. number of people that user $A$ is following, to simulate an equal probability of $A$ to advertise content of any climate-related accounts among its contacts), and multiplied by the importance of $C_1$ in $B$'s profile (to model the likelihood that a user would be exposed to topic $C_1$ when landing on $B$'s page). Formally: $w(T_1,C_1)_{A,B} = \frac{L(T_1)_A \cdot p(C_1)_B }{degree(A)}$, where $p(C_1)_B$ is the percentage of $B$'s videos related to topic $C_1$ (see example in Figure~\ref{fig:q2schema}b). The final weight for each topic pair is the sum of weights across all user pairs.

\begin{figure}[h]
    \centering
\includegraphics[width=0.92\linewidth]{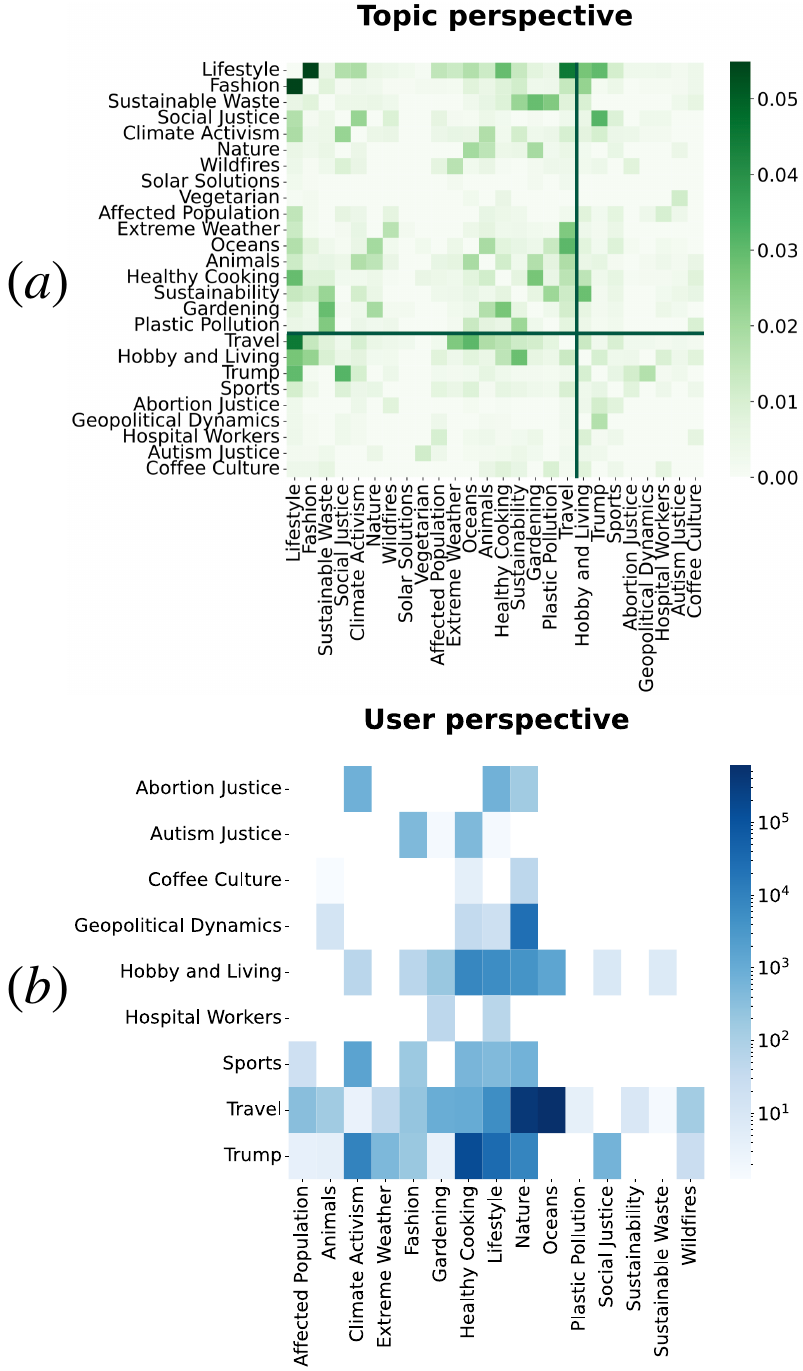}
    \caption{\textbf{Weights on topic pairs.} (a) \textit{Topic perspective.} The heatmap reflects the proportion of shared userbase. (b) \textit{User perspective.} Rows correspond to non-climate topics while columns collect climate-related topics. The weights represent the potential visibility that non-climate topics could bring to climate-related topics.}
    \label{fig:heatmap_res}
   
\end{figure}

The relationships between topics are shown in Figure~\ref{fig:heatmap_res}. From the topic perspective (Figure~\ref{fig:heatmap_res}a), we observe stronger relationships among climate-related topics, especially among \emph{lifestyle} ones. Other notable high-value pairs include \emph{gardening} with \emph{healthy cooking}, and \emph{social justice} with \emph{climate activism}. Concerning pairs between climate and non-climate topics, \emph{travel} emerges as a significant bridge towards climate-related content, showing substantial overlap with \emph{lifestyle}, \emph{extreme weather}, \emph{oceans}, and \emph{animals}. Additionally, noteworthy patterns emerge around \emph{Trump}-related content, which shows strong co-occurrence with both \emph{lifestyle} topics and \emph{social justice} discussions, indicating the intersection of political discourse with both everyday life and environmental advocacy.

From the user perspective (Figure~\ref{fig:heatmap_res}b), we see that influential creators focusing on \emph{travel}, \emph{Trump}, and \emph{hobby and living}-related content show the highest potential for directing audience attention toward climate topics. Among these, \emph{travel}-focused creators demonstrate the strongest potential for engaging their audiences with climate-related content, particularly with topics that share semantic proximity to travel, such as \emph{oceans}, \emph{nature}, \emph{animals}, and \emph{healthy cooking}. \emph{Hobby and living}-oriented creators show promising pathways for influencing engagement with climate-related content such as \emph{healthy cooking}, \emph{lifestyle}, \emph{nature} and \emph{oceans}. Finally, creators focusing on \emph{Trump}-related content exhibit unexpected potential for directing audiences toward \emph{healthy cooking}, \emph{lifestyle} and \emph{nature} topics, as well as \emph{climate activism} and \emph{social justice}.

\section{Discussion and Conclusion} \label{sec:results}

The scope of this study is limited by the constraints that the TikTok APIs put on the volume of information that we could gather during three months of data collection. Additionally, our taxonomy is guided by empirical studies, mainly conducted on Twitter, which together with the manual categorization phase could have influenced the final classification structure; future work could expand and refine our taxonomy. Nevertheless, our work has produced the first public dataset to study the climate discourse on TikTok, while advancing our understanding of climate communication on the platform in two ways. First, we provide a comprehensive mapping of climate-related topics grounded in previous works, and complement it with empirical data to estimate the public engagement on these topics on TikTok. Surprisingly, we find that many topics that are relevant to climate action are not prominently featured on the platform. Second, we identify ``gateway'' topics (e.g., certain themes on politics and travel) with the potential for engaging users who might not primarily seek climate-related content, therefore broadening the reach of climate communicators. This work opens several promising avenues for future research, including the study of the coordination patterns among content creators, and the analysis of the potential impact of endorsement campaigns by non-climate influencers on climate-related topic engagement. Such studies could provide valuable insights into effective strategies for expanding climate awareness and engagement across diverse audience segments on social media platforms.
Importantly, our findings can serve as practical guidelines for campaigns aimed at engaging young audiences in topics related to climate change and climate action.

\section{Related Work} \label{sec:related}
While TikTok has emerged as an influential social media platform, academic research examining its content remains limited, with most previous work on this topic focusing on Twitter~\cite{dahal2019topic, grouverman2019climate}. Existing studies on TikTok have primarily focused on content analysis across select domains, including educational content~\cite{o2023using}, autism awareness~\cite{gilmore2024building}, and preliminary investigations into climate change communication~\cite{basch2022climate, hautea2021showing, nguyen2023tiktok}. Some studies have looked at TikTok usage through surveys~\cite{omar2020watch}.
\citet{basch2022climate} conducted one of the first analyses of videos tagged with \texttt{$\#$climatechange} on TikTok, while~\cite{pera2024shifting}) provided early comparative insights through a cross-platform analysis of climate discourse on TikTok and YouTube on a small set of creators. In this work we looked at the climate discussion from the semantic perspective, using topic modeling. However, previous studies have approached the study of climate discourse from alternative angles including arguments for climate delay~\cite{lamb2020discourses}, narratives around climate change~\cite{flottum2017narratives}, and historical perspectives on the opinions and goals around climate and the climate transition~\cite{rahman2013climate}.

\clearpage

\section*{Ethical Statement}
This study engages with user-generated content on TikTok, where content moderation is platform-dependent. While our analysis maps the landscape of climate change discourse, we acknowledge that some content may express views in ways that could be problematic or counterproductive to constructive collective action. Our broad analytical approach focuses on identifying and understanding the topical structure of climate-related discussions, rather than evaluating individual content for accuracy or potential harm. We recognize the inherent challenges of studying platforms with inconsistent content regulation and the importance of considering their ethical implications.
Furthermore, we acknowledge that the methodological framework presented here could potentially be repurposed for harmful objectives, such as the strategic dissemination of misinformation. While mindful of these risks, we maintain that the potential benefits of this research for addressing climate change challenges outweigh these concerns. Nevertheless, we are dedicated to promoting responsible implementation of our methods/dataset and actively discouraging their misuse.

\section*{Acknowledgments}
We acknowledge the support from the Carlsberg Foundation through the COCOONS project (CF21-0432).

%%%%%%% REFERENCES %%%%%%%
\bibliography{aaai25}

\begin{thebibliography}{35}
\providecommand{\natexlab}[1]{#1}

\bibitem[{Basch, Yalamanchili, and Fera(2022)}]{basch2022climate}
Basch, C.~H.; Yalamanchili, B.; and Fera, J. 2022.
\newblock \#{Climate} {Change} on {TikTok}: {A} {Content} {Analysis} of {Videos}.
\newblock \emph{Journal of Community Health}, 47(1): 163--167.

\bibitem[{Campello, Moulavi, and Sander(2013)}]{campello201hdbscan}
Campello, R.~J.; Moulavi, D.; and Sander, J. 2013.
\newblock Density-based clustering based on hierarchical density estimates.
\newblock In \emph{Pacific-Asia conference on knowledge discovery and data mining}, 160--172. Springer.

\bibitem[{Corso, Pierri, and De~Francisci~Morales(2024)}]{corso2024we}
Corso, F.; Pierri, F.; and De~Francisci~Morales, G. 2024.
\newblock What we can learn from TikTok through its Research API.
\newblock In \emph{16th ACM Web Science Conference}, 110--114.

\bibitem[{Dahal, Kumar, and Li(2019)}]{dahal2019topic}
Dahal, B.; Kumar, S. A.~P.; and Li, Z. 2019.
\newblock Topic modeling and sentiment analysis of global climate change tweets.
\newblock \emph{Social Network Analysis and Mining}, 9(1): 24.

\bibitem[{EcoTokCollective(2024)}]{ecoktokcollective}
EcoTokCollective. 2024.
\newblock EcoTokCollective - Our Team.
\newblock \url{https://www.ecotokcollective.com/our-team}.

\bibitem[{Effrosynidis et~al.(2022)Effrosynidis, Karasakalidis, Sylaios, and Arampatzis}]{effrosynidis2022climate}
Effrosynidis, D.; Karasakalidis, A.~I.; Sylaios, G.; and Arampatzis, A. 2022.
\newblock The climate change Twitter dataset.
\newblock \emph{Expert Systems with Applications}, 204: 117541.

\bibitem[{Falkenberg et~al.(2022)}]{falkenberg2022growing}
Falkenberg, M.; et~al. 2022.
\newblock Growing polarization around climate change on social media.
\newblock \emph{Nature Climate Change}, 12(12): 1114--1121.

\bibitem[{Fl{\o}ttum and Gjerstad(2017)}]{flottum2017narratives}
Fl{\o}ttum, K.; and Gjerstad, {\O}. 2017.
\newblock Narratives in climate change discourse.
\newblock \emph{Wiley Interdisciplinary Reviews: Climate Change}, 8(1): e429.

\bibitem[{Fownes, Yu, and Margolin(2018)}]{fownes2018twitter}
Fownes, J.~R.; Yu, C.; and Margolin, D.~B. 2018.
\newblock Twitter and climate change.
\newblock \emph{Sociology Compass}, 12(6): e12587.

\bibitem[{Gilmore et~al.(2024)Gilmore, Radford, Haas, Shields, Bishop, and Hand}]{gilmore2024building}
Gilmore, D.; Radford, D.; Haas, M.~K.; Shields, M.; Bishop, L.; and Hand, B. 2024.
\newblock Building community and identity online: a content analysis of highly viewed\# autism TikTok videos.
\newblock \emph{Autism in Adulthood}, 6(1): 95--105.

\bibitem[{Goodman(1961)}]{goodman1961snowball}
Goodman, L.~A. 1961.
\newblock Snowball sampling.
\newblock \emph{The annals of mathematical statistics}, 148--170.

\bibitem[{GreenPeace(2023)}]{greenpeaceClimateVoices}
GreenPeace. 2023.
\newblock {C}limate {V}oices on {S}ocial {M}edia of 2023.
\newblock \url{https://www.greenpeace.org/international/story/64559/climate-voices-on-social-media-2023/}.

\bibitem[{Grootendorst(2022)}]{grootendorst2022bertopic}
Grootendorst, M. 2022.
\newblock BERTopic: Neural topic modeling with a class-based TF-IDF procedure.
\newblock \emph{arXiv preprint arXiv:2203.05794}.

\bibitem[{Grouverman et~al.(2019)Grouverman, Kollanyi, Howard, Barash, and Lederer}]{grouverman2019climate}
Grouverman, A.; Kollanyi, B.; Howard, P.; Barash, V.; and Lederer, T. 2019.
\newblock Climate change consensus and skepticism: mapping climate change dialogue on Twitter and Facebook.

\bibitem[{HarvardPique(2024)}]{climatecreatorstowatchClimateCreators}
HarvardPique. 2024.
\newblock {C}limate {C}reators to {W}atch 2024 {W}ebsite.
\newblock \url{https://climatecreatorstowatch.com/}.

\bibitem[{Hautea et~al.(2021)Hautea, Parks, Takahashi, and Zeng}]{hautea2021showing}
Hautea, S.; Parks, P.; Takahashi, B.; and Zeng, J. 2021.
\newblock Showing they care (or don’t): Affective publics and ambivalent climate activism on TikTok.
\newblock \emph{Social media+ society}, 7(2): 20563051211012344.

\bibitem[{Lamb et~al.(2020)Lamb, Mattioli, Levi, Roberts, Capstick, Creutzig, Minx, M{\"u}ller-Hansen, Culhane, and Steinberger}]{lamb2020discourses}
Lamb, W.~F.; Mattioli, G.; Levi, S.; Roberts, J.~T.; Capstick, S.; Creutzig, F.; Minx, J.~C.; M{\"u}ller-Hansen, F.; Culhane, T.; and Steinberger, J.~K. 2020.
\newblock Discourses of climate delay.
\newblock \emph{Global Sustainability}, 3: e17.

\bibitem[{McInnes, Healy, and Melville(2018)}]{mcinnes2018umap}
McInnes, L.; Healy, J.; and Melville, J. 2018.
\newblock Umap: Uniform manifold approximation and projection for dimension reduction.
\newblock \emph{arXiv preprint arXiv:1802.03426}.

\bibitem[{Moulavi et~al.(2014)Moulavi, Jaskowiak, Campello, Zimek, and Sander}]{moulavi2014density}
Moulavi, D.; Jaskowiak, P.~A.; Campello, R.~J.; Zimek, A.; and Sander, J. 2014.
\newblock Density-based clustering validation.
\newblock In \emph{SIAM international conference on data mining}, 839--847.

\bibitem[{Nguyen(2023)}]{nguyen2023tiktok}
Nguyen, H. 2023.
\newblock TikTok as learning analytics data: Framing climate change and data practices.
\newblock In \emph{LAK23: 13th International Learning Analytics and Knowledge Conference}.

\bibitem[{NYTimes(2023)}]{nytimesWithTikTok}
NYTimes. 2023.
\newblock {W}ith {T}ik{T}ok and {L}awsuits, {G}en {Z} {T}akes on {C}limate {C}hange.
\newblock \url{https://www.nytimes.com/2023/08/19/climate/young-climate-activists.html}.

\bibitem[{Omar and Dequan(2020)}]{omar2020watch}
Omar, B.; and Dequan, W. 2020.
\newblock Watch, share or create: The influence of personality traits and user motivation on TikTok mobile video usage.

\bibitem[{OpenAIResearch(2024)}]{OpenAmodels}
OpenAIResearch. 2024.
\newblock OpenAI Terms of Use.
\newblock \url{https://openai.com/policies/usage-policies/}.

\bibitem[{O’Donnell, Jerin, and Mu(2023)}]{o2023using}
O’Donnell, N.; Jerin, S.~I.; and Mu, D. 2023.
\newblock Using TikTok to educate, influence, or inspire? A content analysis of health-related EduTok videos.
\newblock \emph{Journal of Health Communication}, 28(8): 539--551.

\bibitem[{PalauProject(2022)}]{palauprojectEnvironmentalInfluencers}
PalauProject. 2022.
\newblock {E}nvironmental {I}nfluencers {U}sing {T}ik{T}ok {T}o {F}ight {C}limate {C}hange 2022.
\newblock \url{https://www.palauproject.com/blog-post/environmental-influencers-tiktok-climate-change}.

\bibitem[{Pearce et~al.(2019)Pearce, Niederer, {\"O}zkula, and S{\'a}nchez~Querub{\'\i}n}]{pearce2019social}
Pearce, W.; Niederer, S.; {\"O}zkula, S.~M.; and S{\'a}nchez~Querub{\'\i}n, N. 2019.
\newblock The social media life of climate change: Platforms, publics, and future imaginaries.
\newblock \emph{Wiley interdisciplinary reviews: Climate change}.

\bibitem[{Pera and Aiello(2024)}]{pera2024shifting}
Pera, A.; and Aiello, L.~M. 2024.
\newblock Shifting {Climates}: {Climate} {Change} {Communication} from {YouTube} to {TikTok}.
\newblock In \emph{{ACM} {Web} {Science} {Conference}}, 376--381. Stuttgart Germany: ACM.
\newblock ISBN 9798400703348.

\bibitem[{Pupneja et~al.(2023)Pupneja, Zou, Lévy, and Huang}]{pupneja2023understanding}
Pupneja, Y.; Zou, J.; Lévy, S.; and Huang, S. 2023.
\newblock Understanding {Opinions} {Towards} {Climate} {Change} on {Social} {Media}.
\newblock ArXiv:2312.01217.

\bibitem[{Rahman(2013)}]{rahman2013climate}
Rahman, M. I.-u. 2013.
\newblock Climate change: A theoretical review.
\newblock \emph{Interdisciplinary Description of Complex Systems: INDECS}, 11(1): 1--13.

\bibitem[{Reimers(2019)}]{reimers2019sentencebert}
Reimers, N. 2019.
\newblock Sentence-BERT: Sentence Embeddings using Siamese BERT-Networks.
\newblock \emph{arXiv preprint arXiv:1908.10084}.

\bibitem[{Shapiro and Park(2018)}]{shapiro2018climate}
Shapiro, M.~A.; and Park, H.~W. 2018.
\newblock Climate change and YouTube: Deliberation potential in post-video discussions.
\newblock \emph{Environmental Communication}, 12(1): 115--131.

\bibitem[{TikTok(2025)}]{TikTokDevelopers}
TikTok. 2025.
\newblock {T}ik{T}ok for {D}evelopers.
\newblock \url{https://developers.tiktok.com/doc/overview}.

\bibitem[{Treen et~al.(2022)Treen, Williams, O’Neill, and Coan}]{treen2022discussion}
Treen, K.; Williams, H.; O’Neill, S.; and Coan, T.~G. 2022.
\newblock Discussion of climate change on Reddit: Polarized discourse or deliberative debate?
\newblock \emph{Environmental Communication}, 16(5): 680--698.

\bibitem[{Uldam and Askanius(2013)}]{uldam2013online}
Uldam, J.; and Askanius, T. 2013.
\newblock Online civic cultures? Debating climate change activism on YouTube.
\newblock \emph{International Journal of Communication}, 7: 20.

\bibitem[{Wired(2021)}]{wiredMeetClimate}
Wired. 2021.
\newblock {M}eet the {C}limate {C}hange {A}ctivists of {T}ik{T}ok.
\newblock \url{https://www.wired.com/story/climate-change-tiktok-science-communication/}.

\end{thebibliography}

\section*{Paper Checklist}
\begin{itemize}
    
\item For most authors...
\begin{enumerate}
    \item  Would answering this research question advance science without violating social contracts, such as violating privacy norms, perpetuating unfair profiling, exacerbating the socio-economic divide, or implying disrespect to societies or cultures?
    \answerYes{Yes.}
  \item Do your main claims in the abstract and introduction accurately reflect the paper's contributions and scope?
    \answerYes{Yes.}
   \item Do you clarify how the proposed methodological approach is appropriate for the claims made? 
    \answerYes{Yes.}
   \item Do you clarify what are possible artifacts in the data used, given population-specific distributions?
    \answerYes{Yes, we discuss potential biases of our data in the ``Discussion and conclusion'' section.}
  \item Did you describe the limitations of your work?
    \answerYes{Yes, limitations are presented and discussed in the ``Discussion and conclusion'' section.}
  \item Did you discuss any potential negative societal impacts of your work?
    \answerYes{Yes, we address negative societal impact in the ``Ethical considerations'' section.}
      \item Did you discuss any potential misuse of your work?
    \answerYes{Yes, we discuss potential misuse in the ```Ethical considerations'' section.}
    \item Did you describe steps taken to prevent or mitigate potential negative outcomes of the research, such as data and model documentation, data anonymization, responsible release, access control, and the reproducibility of findings?
    \answerYes{Yes, the anonymized data is shared and documented in the anonymised url.}
    
  \item Have you read the ethics review guidelines and ensured that your paper conforms to them?
    \answerYes{Yes.}
\end{enumerate}

\item Additionally, if your study involves hypotheses testing...
\begin{enumerate}
  \item Did you clearly state the assumptions underlying all theoretical results?
    \answerYes{Yes.}
  \item Have you provided justifications for all theoretical results?
    \answerYes{Yes.}
  \item Did you discuss competing hypotheses or theories that might challenge or complement your theoretical results?
    \answerNA{NA}
  \item Have you considered alternative mechanisms or explanations that might account for the same outcomes observed in your study?
    \answerYes{Yes.}
  \item Did you address potential biases or limitations in your theoretical framework?
    \answerYes{Yes, these are addressed in the ``Discussion and conclusion" section.}
  \item Have you related your theoretical results to the existing literature in social science?
    \answerYes{Yes, results are related to existing literature in two sections:   ``Climate Topics on TikTok" and ``Related Works". }
  \item Did you discuss the implications of your theoretical results for policy, practice, or further research in the social science domain?
    \answerYes{Yes, it is discussed in the ``Discussion and conclusion'' section. }
\end{enumerate}

\item Additionally, if you are including theoretical proofs...
\begin{enumerate}
  \item Did you state the full set of assumptions of all theoretical results?
    \answerNA{NA.}
	\item Did you include complete proofs of all theoretical results?
    \answerNA{NA.}
\end{enumerate}

\item Additionally, if you ran machine learning experiments...
\begin{enumerate}
  \item Did you include the code, data, and instructions needed to reproduce the main experimental results (either in the supplemental material or as a URL)?
    \answerYes{Yes.}
  \item Did you specify all the training details (e.g., data splits, hyperparameters, how they were chosen)?
    \answerYes{Yes, HDBSCAN clustering details are reported in the ``Climate Topics on TikTok" section.}
     \item Did you report error bars (e.g., with respect to the random seed after running experiments multiple times)?
    \answerNA{NA}
    \item Did you include the total amount of compute and the type of resources used (e.g., type of GPUs, internal cluster, or cloud provider)?
    \answerNA{NA}
     \item Do you justify how the proposed evaluation is sufficient and appropriate to the claims made? 
    \answerYes{Yes.}
     \item Do you discuss what is ``the cost`` of misclassification and fault (in)tolerance?
    \answerYes{Yes, discussed in the ``Climate Topics on TikTok" section.}
  
\end{enumerate}

\item Additionally, if you are using existing assets (e.g., code, data, models) or curating/releasing new assets, \textbf{without compromising anonymity}...
\begin{enumerate}
  \item If your work uses existing assets, did you cite the creators?
    \answerYes{Yes, we used the tiktok and OpenAI API.}
  \item Did you mention the license of the assets?
    \answerYes{Yes. we provide the links to the TikTok terms in ``Data Collection" and the ones by OpenAI in ``Climate Topics on TikTok" }
  \item Did you include any new assets in the supplemental material or as a URL?
    \answerYes{Yes, anonym.}
  \item Did you discuss whether and how consent was obtained from people whose data you're using/curating?
    \answerNA{NA.}
  \item Did you discuss whether the data you are using/curating contains personally identifiable information or offensive content?
    \answerNA{NA, the identity of creators has no role in the study, as explained in the Datasheet}
\item If you are curating or releasing new datasets, did you discuss how you intend to make your datasets FAIR (see \citet{fair})?
\answerYes{Yes, as cited in the ``Ethical considerations'' section.}
\item If you are curating or releasing new datasets, did you create a Datasheet for the Dataset (see \citet{gebru2021datasheets})? 
\answerYes{Yes, in the GitHub repository.}
\end{enumerate}

\item Additionally, if you used crowdsourcing or conducted research with human subjects, \textbf{without compromising anonymity}...
\begin{enumerate}
  \item Did you include the full text of instructions given to participants and screenshots?
    \answerNA{NA.}
  \item Did you describe any potential participant risks, with mentions of Institutional Review Board (IRB) approvals?
    \answerNA{NA.}
  \item Did you include the estimated hourly wage paid to participants and the total amount spent on participant compensation?
    \answerNA{NA.}
   \item Did you discuss how data is stored, shared, and deidentified?
   \answerNA{NA.}
\end{enumerate}

\end{itemize}
\clearpage

\onecolumn

\section*{Appendix}

\subsection*{Hashtags for video collection }
\begin{table*}[h!]
    \centering
    \begin{tabular}{|c|c|l|}
    \hline
        \textbf{TOPIC} & \textbf{N. HASHTAGS}&  \textbf{ HASHTAGS LIST}\\
        \hline
        \hline
        
ACTIVISM & 6 &  climateaction,   greenwashing,   earthmonth,   savetheearth,   genzforchange,   earthday  \\

CLIMATE & 9 &  climatechange,   climatecrisis,   environmentalist,   climate,   globalwarming,   climativity ,\\&&  eco,   environment,   climatesolutions  \\

DIET & 13 & plantbased,   recipe,   recipes,   food,   foodie,   homemadekitchenrestock,   easyrecipes,   vegan,  \\&& cooking,   kitchenrestock,   foodtiktok,   vegetarian,   sourdough  \\

DISASTER & 32 &  climateemergency,   hurricane,   weather,   tornado,   floods,   tornadoes,   storm,   cyclone,   flood,\\&&   rain,   flooding,   disaster,   naturaldisaster,   fire,   extremeweather,   wildfire,   storms,   severeweather, \\&&  wild,   heatwave,   earthquake,   stormchasing,   extreme,   tornadowarning,   hurricanes,   heatwaves,\\&&   hurricaneseason,   emergency,   water,   thunderstorm,   blizzard,   disasters  \\

ENERGY & 6 & stopwillow,   cleanenergy,   renewableenergy,   fossilfuels,   pollution,   bigoil  \\

HEALTH & 8 &  goodnews,   livebrightly,   health,   spreadlove,   hope core,   climateoptimism,   goodclimatenews, \\&&  todaysgoodnews  \\

JUSTICE & 20 & community,   climatejustice,   environmentaljustice,   indigenous,   nativetiktok,   indigenoustiktok,   \\&& native,  socialjustice,   nativeamerican,   justice,   humanrights,   brown,   browntiktok,   muslims, \\&&   indigenousrights,   environmentalracism,   navajo,   blacktiktok,   muslimtiktok,   racialjustice  \\

LIFESTYLE & 37 & sustainability,   sustainable,   sustainableliving,   ecotok,   ecohome,   minimalist,   slowliving,   thrifted,  \\&&  sustainablestyle,   diy,   sustainablelifestyle,   sustainabilitytiktok,   fastfashion,   travel,   vintage,   thrift,   \\&& thriftedstyle,   environmentallyfriendly,   ecofriendly,   sustainablefashion,   sustainabletravel,   \\&& ecofriendlyliving,   minimalism,   consciousconsumer,   slowfashion,   secondhand,   thrifting,  \\&&  homemade,   gardening,   garden,   sundayreset,   ecotravel,   vanlife,   sustainableswaps,   thriftstorefinds, \\&&   ecofriendlyproducts,   simpleliving \\

 NATURE & 17 & nature,   beach,   ocean,   oceanconservation,   animals,   sand,   earth,   wildlife,   amazon rainforest,  \\&&  marinebiology,   savetheocean,   marinebio,   marinescience,   bees,   marineconservation,   oceanlife,   \\&& conservation  \\

SCIENCE & 8 & learnontiktok,   acespace,   biology,   education,   science,   research,   womeninstem,   moreyouknow \\

 POLITICS & 15 &  cop27,   politics,   cop28,   cop,   democrat,   politicstiktok,   conservative,   cop28dubai,   liberal,\\&&    democrats,   republicans,   political,   politicaltiktok,   cop15,   cop28\_uae \\
 
 WASTE & 22 & lowwaste,   plasticpollution,   trash,   litter,   compost,   plastic,   upcycle,   reuse,   upcycling, \\&&   bandisposablevapes,   foodwastetip,   beachcleanup,   zerowaste,   plasticfree,   cleanup, \\&&  foodwastesolution,    recycle,   foodwasteprevention,   trashboy,   recycling,   reusable,   zerowasteliving  \\
\hline
    \end{tabular}
    \caption{Hashtag list used for the video collection, with the relative macrotopic.}
    \label{tab:topicsize}
\end{table*}

\subsection*{Prompts}

\textbf{Prompt 1: }``\texttt{I have a topic that contains the following document: \textbf{doc}. Based on the above information, can you give a short label of the topic (max 2 words)? Make sure you to only return the label and nothing more.}" \\
where \texttt{\textbf{doc}} is one single video, specifically the concatenation of description and hashtag list after cleaning (stopwords and punctuation removal, filtered out words not in the English corpus) \\
\textbf{Prompt 2:} "\texttt{I have topic that is described by the following keywords: \textbf{subtopics}.Based on the above information, can you give a short label of the topic (max 2 words)? Make sure you to only return the label and nothing more.}". \\
where \texttt{\textbf{subtopics}} refers to the concatenation of all subtopic extracted for the sampled document in the cluster.

\end{document}